\documentclass[sigconf]{acmart}

\usepackage{graphicx}
\usepackage{amssymb}
\usepackage{amsmath}
\usepackage{eurosym}
\usepackage{enumerate}

\setcopyright{none}

\renewcommand\footnotetextcopyrightpermission[1]{} 
\begin{document}
\title{A Blockchain-based Approach for Data Accountability and Provenance Tracking}

\author{Ricardo Neisse, Gary Steri, and Igor Nai-Fovino}
\affiliation{
  \institution{European Commission Joint Research Centre (JRC)}
  \streetaddress{Via E. Fermi 2749}
  \city{Ispra (VA)} 
  \state{Italy} 
  \postcode{I-21027}
}
\email{{ricardo.neisse, gary.steri, igor.nai-fovino}@ec.europa.eu}

\begin{abstract}
The recent approval of the General Data Protection Regulation (GDPR) imposes new data protection requirements on data controllers and processors with respect to the processing of European Union (EU) residents' data.
These requirements consist of a single set of rules that have binding legal status and should be enforced in all EU member states.
In light of these requirements, we propose in this paper the use of a blockchain-based approach to support data accountability and provenance tracking.
Our approach relies on the use of publicly auditable contracts deployed in a blockchain that increase the transparency with respect to the access and usage of data.
We identify and discuss three different models for our approach with different granularity and scalability requirements where contracts can be used to encode data usage policies and provenance tracking information in a privacy-friendly way.
From these three models we designed, implemented, and evaluated a model where contracts are deployed by data subjects for each data controller, and a model where subjects join contracts deployed by data controllers in case they accept the data handling conditions.
Our implementations show in practice the feasibility and limitations of contracts for the purposes identified in this paper.
\end{abstract}


\maketitle

\section{Introduction}
\label{introduction}

The recent approval of the new General Data Protection Regulation (GDPR) \cite{eu:gdpr} in 2016 by the European Commission (EC) imposes new obligations on data controllers and processors in contrast to the previously adopted Data Protection Directive (DPR) \cite{EU1995}.
In a nutshell, the key changes in the scope of the European Union (EU) data protection law introduced by the GDPR in contrast to the DPR that are relevant considering  the contributions of this paper are: it applies to organizations based outside the EU that process personal data of EU residents, it proposes a single set of rules for all EU member states, it extends the responsibility and accountability requirements of organizations, it requires explicit consent with the possibility of withdraw, and it replaces the right to be forgotten by a more limited right to erasure.

The main roles defined in the context of the GDPR are a data subject (person) that authorizes a data controller (organization) to access his/her personal data, with the possibility of forwarding them to a data processor (organization) that will be responsible for processing the data on behalf of the controller\footnote{From this point on we refer to these entities in this paper simply as subject, controller, and processor}.
%
It is the task of the controller to determine the purpose and the manner in which the personal data will be processed according to the constraints imposed by the GDPR.
However, the GDPR also considers automatic delegation of further statutory obligations also to processors that operate as service providers for controllers and process subjects' data.

In light of these new requirements introduced by the GDPR, our focus in this paper is the lack of solutions, from a subject point of view, supporting data accountability and provenance tracking when the subject's data is accessed by controllers and possibly forwarded to data processors.
Subjects should be empowered using a trusted and transparent solution in order to be able to track the controllers and processors that have directly or indirectly accessed their personal data, to verify if the data was accessed, used, and transferred without violating their consent, and to withdraw their consent arbitrarily in case they change their mind or in case they misinterpret the conditions for data usage given in their consent.
An important aspect of such a solution is to enable trust and transparency on data accountability and provenance tracking in a privacy friendly way without increasing the exposure of subjects.
%
From a controller point of view our solution provides a way to prove they received the consent from the subjects to collect their data.

In this context, our goal in this paper is to provide a blockchain-based platform for data accountability and provenance tracking.
We analyze the design issues of this platform and show the design and implementation of two different models.
%
In the first model we implemented the subjects express their consent rules by means of usage control policies that are embedded in specific contracts deployed in the blockchain, which are defined for each controller or processor receiving their data.
In the second model instead, each controller expresses its usage control policies in a blockchain contract with an interface allowing users to join or leave the contract, meaning they are giving or withdrawing their consent for each data controller or processor.
These policies, which can be selected beforehand or on request from a library of policy templates, express the conditions for data access, usage, and transfer to data processors.
With respect to user privacy, data accountability, and data tracking granularity each model provides different properties that are discussed throughout the paper.
%
%
Our contribution is the analysis of design choices, implementation, and performance/scalability analysis of these blockchain-based data accountability and provenance tracking solutions.

This paper is organized as follows.
In section \ref{sec:background} we present background information about blockchain technology, data provenance tracking, and a security policy language that are used as building blocks of our solution.
Section \ref{sec:design-choices} discusses design choices and introduces the different models identified by us for data accountability and provenance tracking using blockchain-based contracts.
Section \ref{model-1} and \ref{model-2} describes the two models we have designed, implemented, and evaluated with different cardinalities of smart contracts and granularities of data accountability and provenance tracking for data subjects, controllers, and processors.
In section \ref{related-work} we compare our proposal with related approaches proposing to use blockchain for data usage accountability.
Section \ref{conclusions} summarizes the conclusions and future work topics.

\section{Background}
\label{sec:background}

In this section we present background information about blockchain-based distributed ledgers and contracts, the adopted data provenance tracking model, and the usage control policy language used to encode the data accountability rules in our approach.
These elements are the building blocks of the data accountability and provenance tracking solution we propose in this paper.

\subsection{Blockchain-based Distributed Ledgers Technology}
\label{sec:blockchain-overview}

The solution for data accountability and provenance tracking proposed in this paper relies on a public blockchain-based distributed ledger platform, namely the open source Ethereum Virtual Machine (EVM) \cite{ethereum}\footnote{The Ethereum public blockchain is implemented using the open source EVM, which can also be used independetly to implement other private or public blockchains. Our focus on this paper is on the use of the EVM technology.}.
Public blockchain platforms such as the EVM can be seen as a distributed database of transactions that can be accessed/managed by many people that do not necessarily trust each other and that do not share a common trusted third party.
%
The  goal of the EVM is twofold:
(1) to keep a decentralized ledger of transactions performed using Ether (ETH), which is the Ethereum virtual currency;
(2) to support the decentralized execution of smart contracts\footnote{From this point on we refer to smart contract simply as a contract always meaning a smart contract deployed in an EVM blockchain}, which are applications that can be assured to run exactly as programed.
Therefore, transactions in the EVM can be transfers of the virtual currency (ETH) from one account to another or execution of contract functions.
In contrast to other public blockchain-based platforms such as Bitcoin \cite{bitcoin} that only support mining of the digital currency and transaction management, the EVM also provides a contract functionality.
At the time of writing Bitcoin and Ethereum are respectively the first and second cryptocurrencies with the largest market capitalization \cite{CoinMarketCap}.
%

Users submitting transactions to the EVM hold a private key (512 bits), which is used to generate a public key (160 bits).
The generated public key is then used to generate a 256 bits EVM address that serves as public user identifiers.
EVM addresses are also used to identify the deployed contracts, however, in contrast to users, contracts do not hold a private/public key pair.
The transactions submitted to the EVM are organized in blocks that are transitively chained to each other using a cryptographic hash function, starting from a pre-computed genesis block.
Once a block of transactions is added to the chain of blocks (a.k.a. blockchain) it cannot be modified or removed since the verification of the chain of hashes would not anymore be valid.
The verification and addition of transactions to blockchains based on the proof-of-work paradigm, is done through the so called "mining" process, which is the consensus algorithm establishing the node entitled to add a new block to the chain. In this process, miners have to solve a cryptographic challenge, and the winner (the first miner that solves it) is rewarded with a certain amount of coins (5 ETH in Ethereum) plus the transaction fees included in the block.
The core idea of the challenge is to regulate the creation of new digital currency and also to reward miners for helping verifying and building the blockchain.
%

For what concerns the contract functionality implemented by the EVM, it simply consists of the capability of allowing the verified deployment and execution of stateful turing-complete programs.
When a contract is deployed, it is assigned a random address that is used in future transactions as a reference to invoke the contract functions.
The invocation of functions that change the contract state must be done using paid transactions; however, EVM nodes can also invoke read-only functions without paying transaction costs, for example, to check the current state of the contract.
The execution fees for contract is computed using a unit named \textit{gas} (short for "gasoline"), and a dynamic market exchange value is defined in relation to ETH\footnote{At the time of writing 1 million gas was priced at 0.02 ETH or \euro0.8. Live stats about the Ethereum blockchain including gas price are available at \url{https://ethstats.net}.}.
When deploying or invoking contract functions users can specify a gas limit for the transaction in order to restrict the amount of ETH they will be willing to pay for the transaction.
A gas limit is also set per block added to the chain, which is currently set to around 4 million gas per block, with a mining rate of approximately 10 seconds per block, which for simply transfers of ETH excluding more expensive contract transactions is capable of processing around 15 transactions per second.
This limit is needed in order to address non-deterministic issues in the creation or execution of a function, for example, the non-termination or excessive usage of blockchain resources since all transactions must be executed and verified by the miners.
In addition to storing the state, contracts may also signal events, which are certified and stored in the blockchain but do not cost as much as a contract state variable.
Contracts can be implemented on an online compiler\footnote{\url{https://ethereum.github.io/browser-solidity/}}, which also allows simulated execution to allow fast testing without the need to deploy in the real or test blockchain.

Many challenges need to be addressed in order to enable such virtual platform including synchronization issues, security, and soundness of the distributed protocol.
Most of these challenges have been addressed and it is out of the scope of this paper to discuss the details about the EVM design and implementation.
Our focus is on the use of the EVM as a distributed platform for tracking data provenance and evaluation of usage control policies in a trustworthy and verifiable way using the contract concept.
We restrict ourselves in this section to a high-level overview of the EVM concepts, and we introduce additional details about contracts as needed to understand our proposed solution in the following sections.

\subsection{Data Provenance Tracking Model}

In order to enable data provenance tracking with smart contracts we follow a structured data model including the specification of data types, data instantiations, and data instances.
In our model primitive and composite data types can be specified, where a composite type includes a named list of contained data instantiations.
For example, date and string can be specified as primitive data types while a user identity can be defined as a list of fullname and date of birth instantiations of these data types.
When data instances about a subject is exchanged with controllers the data model should be agreed beforehand using a data modelling language.
For the purposes of this paper we adopt the data modeling approach used in the  Model-based Security Toolkit (SecKit) \cite{Neisse2015}, which also supports the specification of user identities.
The data provenance tracking model is simply a list maintained by subjects where, for each controller, a list of references to the data provided is updated whenever data is transferred.
When the controller forwards the subject's data to a processor, an entry for the processor with the list of data received is created in the data provenance tracking model.
Since our goal is to store this model in a public blockchain, for privacy reasons we obfuscate the references to the data types, instantiations, and instances.
In the following sections we describe in details how our data provenance tracking model is realized for the two possible models we considered.

\subsection{Usage Control Policy Language}
\label{policy-specification}

Data accountability smart contracts specify restrictions on the usage of the subject's data transferred to controllers.
In our approach we specify these restrictions using the security policy language proposed by the Model-based Security Toolkit (SecKit) \cite{Neisse2015}, already  applied also to Internet of Things \cite{Neisse2014} and mobile devices \cite{Neisse2016} domains.
In the SecKit policy language detective and preventive mechanisms are specified using  ECA (Event-Condition-Action) rules.
%
%
Preventive mechanisms allow the specification of rules that do not allow an undesired behavior to occur, while reactive mechanisms simply take compensation actions when something undesired is observed.
In order to support these two types of mechanisms the Event part of the rule considers tentative and actual events, representing respectively activities that are about to take place but were not yet executed or activities already happened.
In this paper we are concerned only with preventive mechanisms since our goal is to disallow undesired behaviors by controllers, for example, to prevent data from being misused or exchanged with 3rd parties.

In the Action part of preventive mechanisms enforcement actions can be specified to allow, deny, modify, or delay the execution of data usage activities by controllers.
In the Condition part a logical expression using a combination of operators can be specified including event pattern matching, trust relationships, context-based, role-based, propositional, temporal, and cardinality operators.
With this policy language it is possible to express, for example, that during a time window (e.g., 30 days) controllers should be allowed to perform an operation only once.
Policies are evaluated using a discrete time step window with a configurable time granularity that is configured according to the policy requirements, for example, if the policy refers to restrictions on data usage considering hours, days, months, etc.
For details about all operators supported in our policy language we refer the reader to \cite{Neisse2015}.

The policy language proposed by the SecKit also supports the specification of modular configurable mechanism templates using variables.
Templates can be dynamically instantiated and disposed using ECA rules following the same mechanism structure.
This is particularly useful in order to enable re-use of mechanisms without repeating patterns of enforcement that should be applied for multiple subjects or situations.
In the description of our approach we show examples of preventive mechanisms and templates encoded in smart contracts and describe how these mechanisms can be evaluated in the blockchain in a privacy-friendly way.

\section{Solution Design Choices}
\label{sec:design-choices}

In this section we describe the three different models for blockchain-based data accountability and provenance tracking we identified.
When designing such a solution, a few decisions should be taken with respect to the design of contracts and the blockchain architecture in order to proper address performance and authorization issues.
The contract design issues are related to their lifecycle, the required state variables to store the contract information, and authorization policies specifying who should be allowed to read and update the contract variables.
The blockchain architecture decisions are concerned with the adoption of a public, consortium/semi-public, or private blockchain solution considering transaction management issues such as authorization and auditability properties.
In the following, we first introduce these three different models and discuss design and architecture issues that should be addressed when implementing these models.
The architecture issues we discussed are relevant only on the real-world deployment of the proposed models; in this paper, we are mostly concerned with design, implementation, and performance of these models using the EVM platform.

With respect to the contract lifecycle, we identified three possible models with different cardinality of contracts in relation to the number of data subjects and controllers: 

\begin{enumerate}[(a)]
	
	\item \textbf{Data subject contract for specific controller}: the subject creates a contract tailored for each controller that manages his/her data. The contract keeps track of the data shared with the controller, the policies regulating the use of the data, and registers the data usage events representing the activities performed by the controller using the subject data as input;
	
	\item \textbf{Data subject contract for specific data}: the subject creates a generic contract for each data instance that is shared for all controllers accessing the data. The contract contains the list of controllers that were given access to a particular data instance and the policies they should respect;
	
	\item \textbf{Controller contract for multiple data subjects}: the controller creates a contract specifying how the data received from all subjects is treated. Data subjects then join the contract in case they accept the data usage policies of the controller, similarly to the Platform for Privacy Preferences Project (P3P) \cite{p3p}.
	
\end{enumerate}

\textbf{Cardinality of contracts and accountability granularity}.
The models differ with respect to the cardinality of contracts, the level of customization, and the data tracking granularity allowed to subjects.
Model (a) has the highest number of contracts since there is a custom tailored subject contract for each controller accessing subject's data, which is more adequate when very sensitive data is exchanged (e.g., health information).
In model (b) the subject creates one contract for each data type that is possibly accessed by many controllers.
In model (c) there is just one contract for each controller, which is expected to be several orders of magnitude lower than the number of subjects, but also has the lower level of customization since subjects are left to choose between a limited number of contract options.
Only model (a) and (b) consider the registration of fine-grained data provenance information including data usage events.

\textbf{Trust model}.
A blockchain-based platform is essentially a public decentralized database without a single point of failure that can be trusted for correctness but not for privacy, can be accessed/managed by many people that do not necessarily trust each other and do not share a common trusted third party.
A public blockchain-based solution is needed in this scenario since the following conditions are met: (1) there is a need for a database tracking data provenance and usage, (2) many organizations need write access to this database, (3) the organizations that write to the database may not necessarily trust each other, and (4) the organizations do not have a shared trusted third party managing the database.
Controllers and/or processors may intent to be legally bound to the EU law but may not trust an EU institution to manage this database since according to the GDPR they may also be organizations outside the EU.
Since the EU regulation applies to subjects that are EU residents but maybe not EU citizens, they might not trust an EU institution for managing this database.

\textbf{Public, consortium, or private blockchain}.
With respect to the blockchain architecture, the adoption of a public, semi-public, or private approach has an impact on the level of public auditability and censorship resistance properties of the solution \cite{VitalikPublicPrivateBlockchains}.
A public blockchain approach such as the Bitcoin and Ethereum public blockchains allows anyone to read, send transactions, and participate in the consensus/mining process to determine which blocks are added to the chain and determine the current blockchain state.
In a semi-public or consortium  blockchain read access may be public or restricted and the consensus process is controlled by a set of organizations.
In a private blockchain read access may be public or restricted and write permissions are decided by one central organization.
The adoption of any of these approaches has an impact on the censorship resistance, privacy, anonymity, performance, and scalability.

\textbf{Censorship resistance}.
Contract creation and joining is on the interest of controllers, since it allows them to receive and process data. However, subjects withdrawing their consent is not on their interest, because the data would be no longer available for processing.
Considering this fact, the adopted blockchain solution must ensure censorship resistance.
In fact, if a private blockchain solution is adopted, the central authority could simply refuse to add transactions where subjects are withdrawing their consent in order to keep the rights to use the subjects data.
Therefore, it is important to make sure transactions are fairly processed considering a minimum quality-of-service requirement, for example, it should be guaranteed that all subject transactions are added to the blockchain at most one day after the transaction submission.
Data protection authorities can assume the role of verifying subject complains and censorship resistance in case a private blockchain solution is adopted.


\textbf{Privacy and anonymity}.
From an anonymity point of view model (b) is not acceptable since subjects will need to use a unique address known by all controllers allowing for identity linkability between controllers.
Models (a) and (c) allow subjects to use different pseudo-identifiers (EVM addresses) and by design prevent direct linkability of subscriptions among different controllers.
In model (a) the only thing that can compromise the privacy of subjects is the policy structure since in our implementation we adopt a privacy-friendly way to encode data provenance and data usage events without revealing the details about the actual data instances exchanged with controllers (see Section \ref{model-1}).

\textbf{Performance and Scalability}.
From a scalability point of view the controller generic approach is the best one, but it also constrains a possible solution to public blockchains due to the high number of transactions.
A service provider such as Facebook has on average 5 new users per second\footnote{https://zephoria.com/top-15-valuable-facebook-statistics/}, which would mean 5 transactions per second only to manage the joining of these new subjects to the data usage contract.
Considering the transaction rate of the current public chains where Bitcoin has a theoretical limit of 7 transactions per second (tx/s) but usually reaches on average only 3 tx/s \footnote{https://blockchain.info/charts/n-transactions} and Ethereum can reach around 15 tx/s, this throughput is clearly not enough with the number of Facebook-like service providers currently being used.
The only feasible solution is an independent blockchain run by the company itself, which can be implemented using a private EVM chain, with a checkpoint on a public chain to verify the transactions.
By running this in a private blockchain operated by the service provider subjects are at risk of not having their opt-out from a privacy policy registered, since it could be just ignored by the provider.
However, we expect this to be solved with a cryptographic acknowledgment of the transaction by the service provider, which allows subjects to proof they handled their request to the blockchain.

\textbf{Economics}.
In case a public blockchain approach is taken the benefits regarding non censorship and public auditability are definitely compelling.
However, in a public blockchain the transaction fees have to be paid for the processing contract invocations, which could make the approach (a) unfeasible due to the high number of transactions.
Subjects and controllers would need to agree on an economic model to device who pays for the contract fees.
This is an additional argument to adopt a private blockchain solution where processing fees could be reduced significantly.
We do not present in this paper a complete economic analysis of our solutions but from our evaluation results discussed in Sections \ref{model-1} and \ref{model-2} and considering the large number of subjects and controllers that process data, it is evident that these models could not be adopted only relying on the public EVM chain.

\textbf{Off-blockchain communication}.
All proposed models consider the need for off-blockchain communication requirements including the transfer of the subject's data to the controllers.
The purpose of the blockchain is to register and provide auditability of the communications when needed.
The EVM proposes Swarm and Whisper respectively for off-blockchain decentralized data storage/distribution and messaging.
The integration of these approaches in our solution is out of the scope of this paper.

\textbf{Guarantees}.
The goal of our approach is not to prevent controller and providers from unlawfully holding or redistributing data, but to enable verifiability of data custody and a mechanism for subjects to issue notifications about the misuse of their data.
From a controller and processor point of view the main benefit is a certified proof that can be presented to supervisory authorities showing the data was obtained in a GDPR compliant way.

\textbf{Data query service}.
Controllers and processors provide an off-blockchain service allowing subjects to query the stored data about them in order to verify the hashes in the blockchain.
This is needed in case of implicit data collection, which subjects are usually unaware of.
This is out of the scope of this paper.

\textbf{Applicability}.
From an applicability point of view we identified two possible scenarios with respect to the GDPR compliance for our models.
First, our solution could enable subjects to grant consent to controllers, specify their data usage control policies, and track the data provenance flow in a trusted and privacy-friendly way.
By observing the data provenance, flow subjects may correct their policies in case an unwanted flow takes place, or even revoke access to the data from that particular point on.
Secondly, it provides a regulatory framework mechanism to enable GDPR compliance checking by Supervisory Authorities in case a non-compliance activity (i.e., an unlawful data access) is signaled by subjects.
For example, when a subject receives an unsolicited e-mail advertisement from an organization containing his/her personal data for which provenance cannot be determined, it could be an indication of a non-compliant activity.
The signaling of non-compliant activities could trigger investigation and auditing of the organization by the Supervisory Authorities and possibly lead to a sanction in case the organization cannot show by informing the respective transactions in the blockchain allowing the data to be received and used for the specific purpose.
Subjects could also by default require all interactions with them to be backed by links to the respective transactions in the blockchain, even in non-digital communications, for instance, paper letters could contain a machine readable code encoding the transaction number allowing the specific data to be collected.
Subjects could then read this code using their mobile phone and check the compliance of the obtained personal data.

\section{Subject Contract Model}
\label{model-1}

\begin{figure}[!h]
	\centering
	\includegraphics[width=\columnwidth]{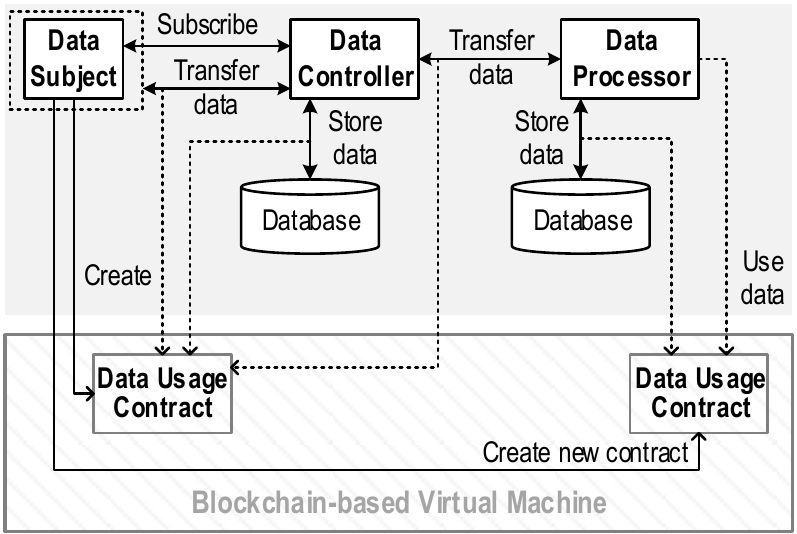}
	\caption{Architecture} 
	\label{fig:arch}
\end{figure}

Figure \ref{fig:arch} presents the high-level architecture of the more fine-grained data accountability and provenance tracking model proposed in this paper.
In this architecture, three main entities are depicted following the GDPR terminology: the \emph{Data Subject}, the \emph{Data Controller}, and the \emph{Data Processor}.

When the subject subscribes with a controller, which is typically the role of a service provider, it creates a policy-based \emph{Data Usage Contract} specifying constraints on the usage and redistribution of any data obtained explicitly or implicitly by the controller.
Explicit data is any data provided directly through interactions with the subject such as the e-mail addresses or birth date.
Implicit data is any data acquired automatically, for example, sensor data from IoT devices in the environment surrounding the subject, data acquired by apps installed in mobile devices, or even server log files registering details of the network interactions between subject and controller services (e.g., IP addresses).
The contract in this model acts as a data provenance tracker, policy evaluation entity, and event logger that allow the subject to easily check all data transfers and usage transactions providing assurance that only transactions conforming to the contract policies are authorized and registered in the blockchain.

\begin{figure}[!h]
	\centering
	\includegraphics[width=\columnwidth]{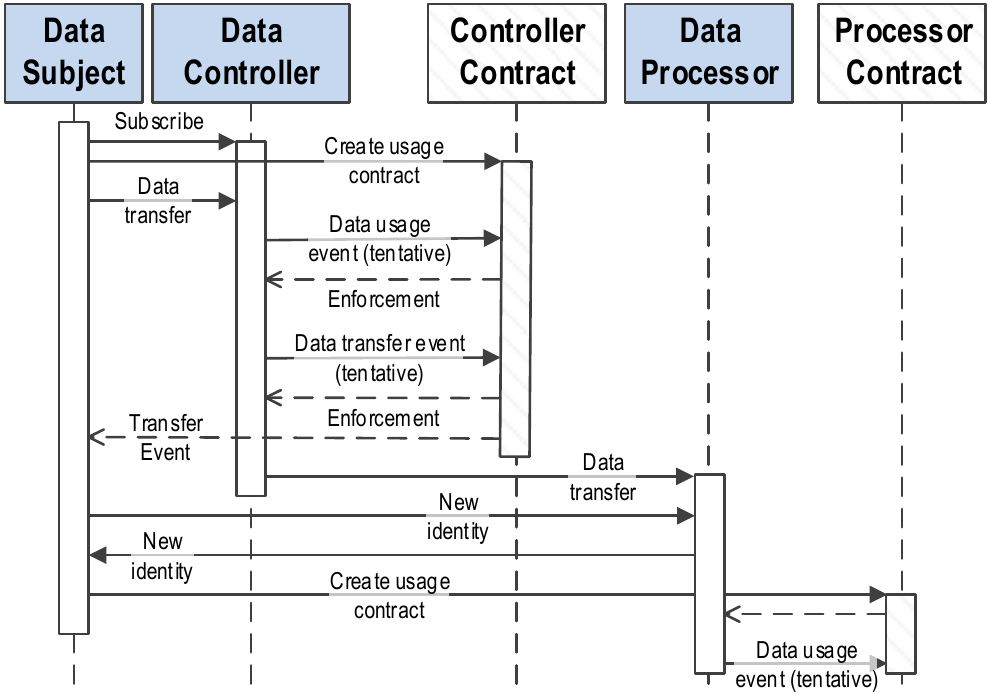}
	\caption{Sequence diagram} 
	\label{fig:seq}
\end{figure}

Figure \ref{fig:seq} shows the sequence diagram of the interactions between the entities and contracts.
%
The subject subscribes with the data controller, creates the custom contract for this controller regulating the use of his/her data, and transfers the data to the controller.
%
For each new established contract the subject uses a new blockchain address to prevent linkability of the contracts established with each controller, requiring the subject to maintain a list of all addresses used and the respective nonce established with each controller or processor.
%
After creating the contract, the subject transfers the data to the controller.

\begin{figure}[!h]
	\centering
	\includegraphics[width=\columnwidth]{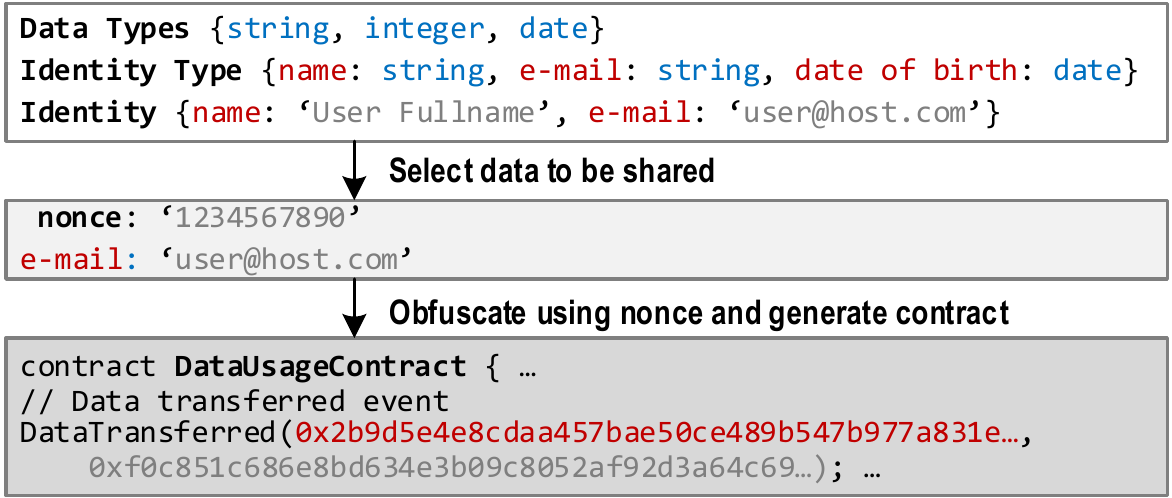}
	\caption{Data provenance encoding in contract} 
	\label{fig:data_provenance}
\end{figure}

The newly created contract stores a list of the data transferred to the controller including data instantiation and instance values.
This information is not store in plain since the contract is deployed in a public blockchain where anyone is able to access the contract state variables and transactions, which would imply public access to all the subject's data.
Figure \ref{fig:data_provenance} shows the schema we adopt to obfuscate the provenance tracking information stored in the contract.
For each controller the subject selects the data to be shared, which is encoded according to a data model, and establishes a secret random nonce that is only shared with the data controller in an encrypted format using the contract initialization parameters.
The subject then stores in the contract only the hashes (generated using the SHA3-256 function) of the data instantiations and data instances values (e.g., e-mail="...") concatenated with the nonce, which is only known at this point by the subject and the controller.
The objective is to provide indistinguishability and to prevent crawling of the blockchain searching for the same data value in order to re-identify the subject contracts, for example, if a controller would know also the same data value (e.g., e-mail) it could simply search the blockchain transactions for this same exact hash.

%
Whenever the controller is about to perform a data usage activity it should check using a read-only local transaction to the contract interface if this activity is allowed according to the policies.
In case the activity is allowed by the contract, the controller can record it using a blockchain transaction signaling the actual data usage event, which is recorded in the blockchain for a later accountability check.
Data usage activities include access to the subject's data, storage of the data in a local database managed by the controller, execution of any purpose specific activity, generation of derived or consolidated data, and transfer/redistribution of data to processor.
In order to protect the privacy of subjects, events and policies are also anonymized using a hash function and the nonce already established with the data controller.
The behavior of the policy is encoded in the contract and uses a privacy-friendly policy evaluation algorithm that also relies on hash functions.

\begin{figure}[!h]
	\centering
	\includegraphics[width=\columnwidth]{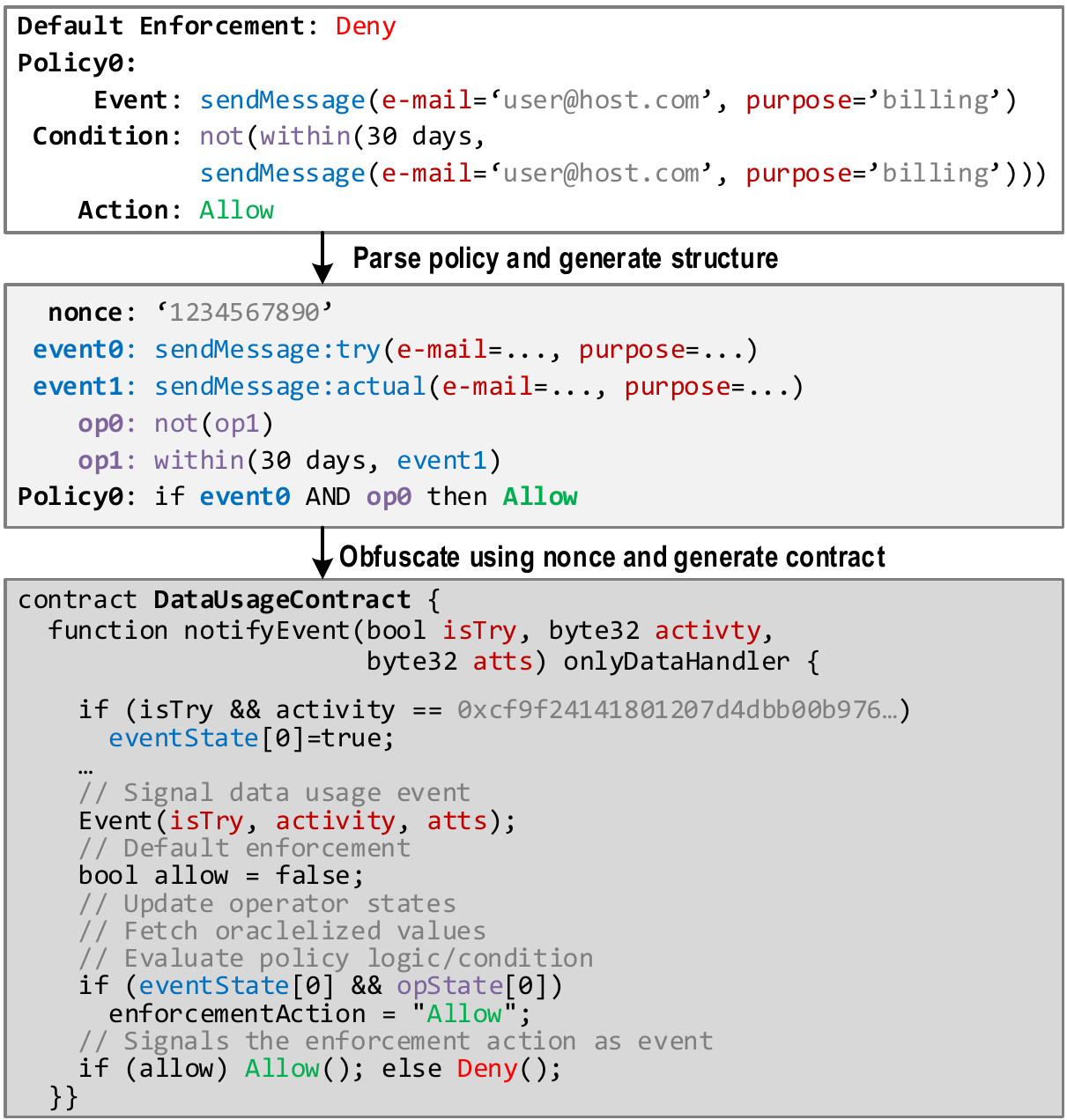}
	\caption{Generation of contract from policy} 
	\label{fig:policy_generation}
\end{figure}

Figure \ref{fig:policy_generation} describes how the policy behavior is encoded in a contract for a sample policy that allows a billing message to be sent to the subject e-mail at most once every 30 days and denies any other usage activities.
%
%
The first step is the parsing of the policy structure and identification of the event patterns and operators creating a tree of the policy structure.
For each event pattern specified in the policy, the activity name and attributes are obfuscated using the secret random nonce and a hash function, in the same way already done for the data provenance information.
The event matching is then encoded in the $notifyEvent$ function of the contract, which updates the states of the operator tree and verifies if the final evaluation of the policy is true or false.
We do not store the events in the contract, we simply update the states in case the event is observed, and for cardinality and temporal operators a time step window with the previously observed events is updated.
The encoding of each operator is custom considering the specified policy, for example, the $within$ operator just requires a boolean flag and the last update time for evaluation purposes.
Reactive policies that are triggered based on time instead of an event such as "raise a notification if my data is not deleted in 30 days" require the subject to verify periodically the blockchain state in order to observe violations and generate a time step event in order to trigger the execution.
In case the user does not generate the time step event the violation is detected whenever the contract is notified about any relevant event.
In our current implementation we only support equality matching of obfuscated events and attributes, in order to support other comparisons such as "allow something if value is bigger or lower then a threshold"  would require more sophisticated schemes such as homomorphic encryption.
Furthermore, if policies depend on any external information (e.g., context or trust relationships), this information has to be pushed in the blockchain\footnote{\url{https://github.com/oraclize/ethereum-api}}.

The activity of transferring data from a controller to a processor should be also authorized by the contract policies.
In case the transfer is authorized by the contract, the data is transferred and an event is generated containing the address of the processor known by the controller in an encrypted format only readable by the subject.
%
The subject is able then to establish a new identity with the processor, which is even unaware of the controller contract address established with the subject.
Finally, the subject creates a data usage contract for the processor following the approach used with the controller; the processor contract consists of a new shared secret nonce, the data provenance information, and the encoded policy behavior that should be enforced by the controller.

Subjects may withdraw their consent by deleting the contract from the blockchain, which simply makes the contract inactive from that moment in time on, while the complete contract history stays forever recorded.
Furthermore, users may also release additional data or impose more policies/constraints on the data released to the controllers/processors.
This can be achieved by adding additional data provenance information to the parent contract and by creating child contracts encoding more desired policy behaviors.

We have implemented a sample contract using the policy example in Figure \ref{fig:policy_generation} and evaluated the performance for transactions creating and executing contract functions in a private blockchain implemented using the EVM.
In our implementation we evaluated two choices regarding the storage of data usage events in the contract state variables.
In the first version 10 hashes of data types and instances were stored as state variables in the contract with a total of 736 bytes (2 times byte32 for each data), while in the second version the data hashes were signaled as events associated to the contract (each event signaling 2 times byte32).
Just creating the contract without any data associated consumed 0.82 million gas, with the events the consumption increased to 0.89 million gas, and with the state variables the consumption was as high as 1.25 million gas.
These results show that it is far better to use events in case of simply logging activities instead of using contract state variables.
We also evaluated the cost of notifying a data usage event to the contract, which generates a contract event with the respective enforcement result (i.e. allow, deny, modify or delay).
The event notification cost was only of 0.023 million gas.

Since contracts have a limited amount of storage and processing capabilities it is unfeasible to allow flexible configuration of policies in a single contract with a more complex interfaces.
Contracts should embed the policy logic and in case a new policy or data transfer takes place a new child contract should be created.
Data controllers are allowed to add new child contracts to the existing one where two approaches can be followed depending on the number of users and constraints.
All users that joined a parent contract are automatically added to the child contract unless they add a restriction.
All users that joined a parent contract must explicitly re-join the child contract to be included in the new service conditions.
Nevertheless users are able to watch the contracts and be notified about these events in order to be able to react.
Data controllers may choose to link new child contracts with additional features that users will be allowed to receive from the service by associating data requirements to the provided features.

With respect to user's privacy, this model guarantees user pseudoymity and unlinkability among different controllers and processors and  unobservability by third party entities of the user policies and data events in the public blockchain that are necessary to evaluate the respective policies.
Linkability can be achieved in case processors and controllers collude by exchanging their addresses and nonces used in each contract if users provide unique identifiers (e.g. e-mail) allowing entities that received this identifier to know the associated addresses of a user.
Our solution also considers that, in particular cases, upon request of a third party such as a the Data Protection Supervisor Authority, both subject or controller/processor may be requested to prove the possession of the data and to prove the data usage activities performed did not violate the subject's policies.
In this case, the release of the encrypted nonce by subject and/or processors/controllers is enough to enable verification of the policies and the data provenance information.

\section{Controller Contract Model}
\label{model-2}

%
%
%
%
%

In this section we describe the more coarse-grained data accountability and provenance tracking model proposed in this paper.
%
%
In this model the controller or processor creates a contract specifying the shared constraints on the usage and redistribution of any explicit or implicit data obtained from all subjects that subscribe with the controller.
The contract in this model acts simply as a repository of the configurable policy templates that will be instantiated for all subjects and do not store anything except the list of subjects.
Transactions are only requests by subjects to join or leave the contract.

\begin{figure}[!h]
	\centering
	\includegraphics[width=\columnwidth]{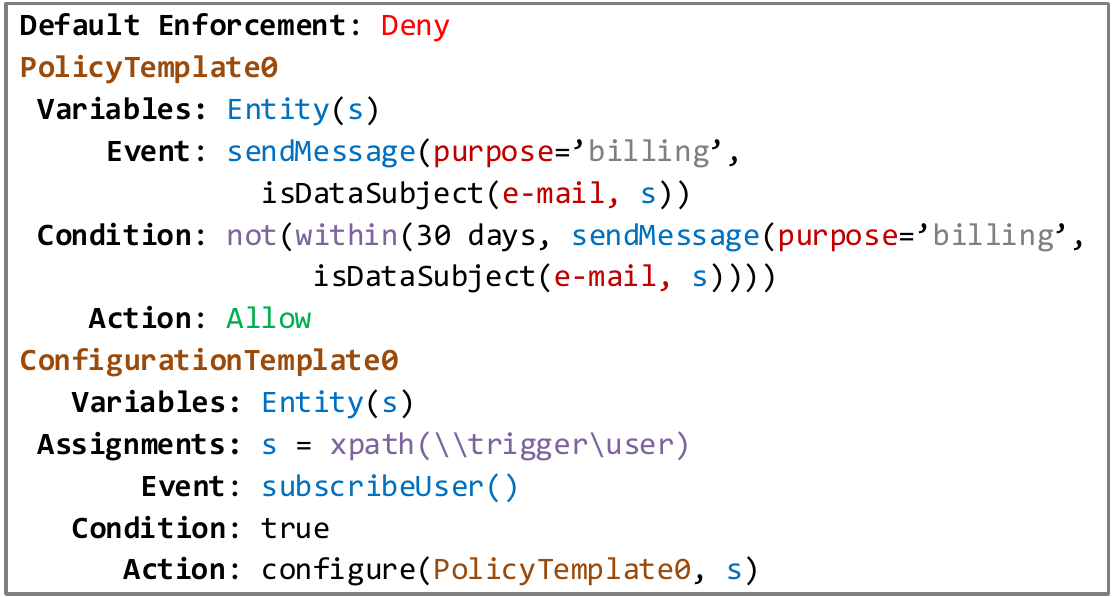}
	\caption{Enforcement and configuration templates} 
	\label{fig:policy1}
\end{figure}

Figure \ref{fig:policy1} shows an example of policy enforcement and configuration templates that can be used in a data usage contract in this model.
The policy enforcement template specifies an enforcement pattern using  an entity variable $s$, which in this example is one variable for the data subject that allows a billing message to be sent no more than once per month.
This policy is configured for all users that subscribe to the service in the defined configuration template, and a special data assignment function is used to determine in the policy evaluation if the e-mail is associated to the data subject encoded in the variable.

\begin{figure}[!h]
	\centering
	\includegraphics[width=\columnwidth]{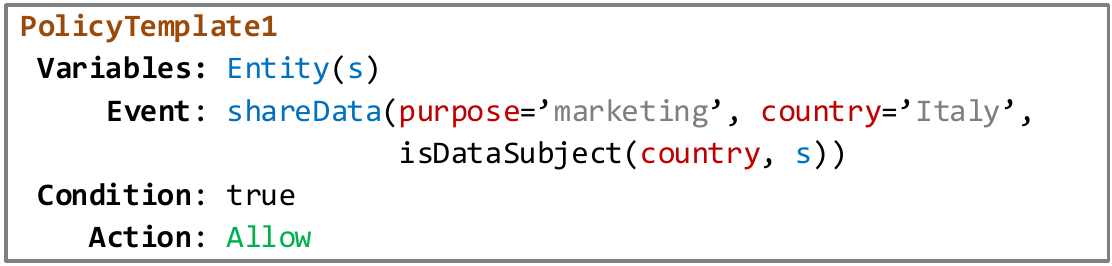}
	\caption{Data sharing policy for subjects of a particular country} 
	\label{fig:policy2}
\end{figure}

Since this model is more focused on controllers processing large amounts of data we also allow the data controller to log events representing bulk executions of the contract, for example, using an event template send message where the e-mail data attribute is left unspecified.
Bulk events make sense when regular activities are completed nearly at the same specific moment in time such as on a daily time window.
Bulk events may also be parametrized, for example, in case an activity happens for a subset of subscribers matching a particular condition using also a policy template.
Bulk attributes may be specified using the same ECA template of policies using a tentative event structure, for example, all data of data subjects from a particular country will be shared.
This example policy is specified in \ref{fig:policy2}.

We have implemented a sample contract where the policy is configured as a string, which could be an XML or JSON encoding of the policy specified using the SecKit policy language.
As expected, this contract requires a much lower amount of gas in contrast to the model detailed in the previous section.
The creation of the contract requires only 0.34 million gas and the joining and leaving of data subjects only requires around 40 thousand gas.
The contract does not requires any other operation since it does not log any data usage events or enforcement actions.

\section{Related Work}
\label{related-work}

The nature of blockchain is particularly suitable for tracking assets exchanged by different entities. In their most famous implementations, actually blockchains do not exchange any asset (i.e. coin or token), but the ownership of a certain amount of them is just the result of evaluating the transaction list containing the quantity of assets exchanged between two entities. Therefore, when applied to digital or physical assets, it is fundamental that the latter can be unequivocally represented and identified, and that they cannot be (trivially) cloned.

One of the first application of blockchains to the tracking of physical goods was for the provenance of diamonds \cite{Everledger}. Here the strong point is that diamonds can be uniquely fingerprinted through their physical properties, making them distinguishable one from another.
This fingerprinting allows the diamond discovery processing and registration in the blockchain to easily match the concept of mining in a standard proof-of-work based blockchain system.

When the assets to be tracked are digital, the first problem is related to the possibility of easily cloning and transferring them, so that the main goal is at least to prove their first origin.
The tracking of digital data provenance using blockchains was first proposed in a discussion paper presented at the ID2020 Design Workshop \cite{ImmutableMe}. Although at high level, the paper describes a proof of concept for verification and attestation of provenance of individuals' data using blockchains. The latter do not directly store the data, which are referenced with hashes and associated with tokens, so that it is possible to prove their (immutable) origin and track all the data exchanges. Although the proposed approach is novel, it does not cover the definition af advanced policies or contracts regulating the usage of the exchanged data in the way proposed by us in this paper.

An example of users' data protection and privacy policy enforcement on a blockchain is provided in \cite{PoliciesOnBitcoin}. The authors propose to control access permissions to private data collected by a service (e.g., location from a mobile phone) through a Bitcoin blockchain.
Every time a users subscribes to a service a new transaction specifies the access permissions and another contains the hash of the data, which are stored in a off-chain database. Policies encoded in a protocol executed by the blockchain grant or deny access to the data referenced in the chain. Although this solution is in part similar to ours, their proposed policies only specify simple allow/deny enforcement (i.e. white/blacklisting) without the possibility to express more complex policy conditions. Moreover, scalability is not taken into consideration: issuing minimum two transactions for every subscription of a user application to an online service would easily saturate a Bitcoin network even considering few service providers.

Inspired by the example above, is the application of data tracking to health care proposed in \cite{HealthCare}. Here the blockchain is again a medium to control the access to data stored off chain. The paper provides only a high level description, but already identifies the main limitations, mainly in terms of scalability and data privacy, of Bitcoin-based blockchains for this kind of applications.

Applied to the same field but with a more comprehensive approach is MedRec \cite{MedRec}. The authors propose to give patients control over their Electronic Health Records (EHRs) through the use of Ethereum blockchain and smart contracts, which manage authentication, confidentiality, accountability and sharing of the data. The latter are referenced in the chain using their hashes but are stored externally. Miners are rewarded with anonymized aggregated data useful for research. Although the proposed approach addresses many issues, considerations about costs and scalability on a large scale deployment are not mentioned.







\section{Conclusions and Future Work}
\label{conclusions}

In this paper we analyze the feasibility of using a blockchain-based contract approach to support data accountability and provenance tracking in light of the new requirements of the GDPR.
We discuss several solution design choices and introduce three different models, including two concrete implementations, with an extensive analysis with respect to data accountability and provenance tracking granularity, privacy, anonymity, performance, and scalability.
We show that for more sensitive data with less frequent exchanges, such as medical data, a more fine-grained solution where subjects create contracts with each controller and processors is more adequate.
On the other hand, for more dynamic data with more frequent exchanges and strict scalability and performance requirements, controllers or processors should manage a contract that registers all subjects accepting all or part of the data usage conditions.
%

A possible solution for scalability issues we are currently investigating is the use of sharding, where the blockchain is divided in separate chains that that are responsible for contracts of a subset of all controllers and processors.
These separate private chains then synchronize with the public chain on regular intervals, for example every 1000 blocks, in order to allow for public verifiability \cite{sharding}.
In case the separated chains are managed privately, data protection supervisory authorities can then join all chains just as observers in order to prevent censorship and guarantee that transactions of data subjects are not indiscriminately refused.

As future work we also plan to investigate the possibility of using business blockchain approaches such as the Hyperledger solution, which uses a different algorithm for reaching consensus and also has a more ambitious scalability and performance goal with thousands of transactions per second \cite{bcc2017,Vukolic2016}.
Hyperledger is the leading business oriented blockchain-based platform that supports a modular consensus protocol and is currently developed and supported by a large consortium of high profile companies such as IBM and Intel.

We also plan to work on a model-based translation mechanism to automatically generate contracts from policies specified in our policy language.
Our approach in this paper was simply a manual translation in order to demonstrate the feasibility and to investigate the strengths and limitations of the technology.
%
%
Finally we plan to deploy and evaluate scalability and performance issues related to the expected size of the blockchain after a number of contracts and events is added to the chain and also to verify the transaction throughput that can be achieved.

\bibliographystyle{ACM-Reference-Format}
\bibliography{all}

\end{document}